**RESEARCH ARTICLE**                                                          **Open Access**

# Modulation of transforming growth factor beta signalling pathway genes by transforming growth factor beta in human osteoarthritic chondrocytes: involvement of Sp1 in both early and late response cells to transforming growth factor beta

Catherine Baugé[1*], Olivier Cauvard[1], Sylvain Leclercq[2], Philippe Galéra[1], Karim Boumédiene[1*]

## Abstract

**Introduction:** Transforming growth factor beta (TGFβ) plays a central role in morphogenesis, growth, and cell differentiation. This cytokine is particularly important in cartilage where it regulates cell proliferation and extracellular matrix synthesis. While the action of TGFβ on chondrocyte metabolism has been extensively catalogued, the modulation of specific genes that function as mediators of TGFβ signalling is poorly defined. In the current study, elements of the Smad component of the TGFβ intracellular signalling system and TGFβ receptors were characterised in human chondrocytes upon TGFβ1 treatment.

**Methods:** Human articular chondrocytes were incubated with TGFβ1. Then, mRNA and protein levels of TGFβ receptors and Smads were analysed by RT-PCR and western blot analysis. The role of specific protein 1 (Sp1) was investigated by gain and loss of function (inhibitor, siRNA, expression vector).

**Results:** We showed that TGFβ1 regulates mRNA levels of its own receptors, and of Smad3 and Smad7. It modulates TGFβ receptors post-transcriptionally by affecting their mRNA stability, but does not change the Smad-3 and Smad-7 mRNA half-life span, suggesting a potential transcriptional effect on these genes. Moreover, the transcriptional factor Sp1, which is downregulated by TGFβ1, is involved in the repression of both TGFβ receptors but not in the modulation of Smad3 and Smad7. Interestingly, Sp1 ectopic expression permitted also to maintain a similar expression pattern to early response to TGFβ at 24 hours of treatment. It restored the induction of Sox9 and COL2A1 and blocked the late response (repression of aggrecan, induction of COL1A1 and COL10A1).

**Conclusions:** These data help to better understand the negative feedback loop in the TGFβ signalling system, and enlighten an interesting role of Sp1 to regulate TGFβ response.

## Introduction

Transforming growth factor beta (TGFβ) controls a wide range of cellular responses, including differentiation, cell proliferation, migration, apoptosis, extracellular matrix remodelling and development. In cartilage, TGFβ plays a crucial role by functioning as a potent regulator of chondrocyte proliferation and differentiation, and of extracellular matrix deposition [1].

Biological effects of TGFβ are mediated by two different serine/threonine kinase receptors, named type I (TβRI) and type II (TβRII), which are both required for inducing signal transduction. Following binding of TGFβ to TβRII, the ligand-bound type II receptor forms an oligomeric complex with the type I receptor, resulting in TβRI phosphorylation. Activated TβRI (also called ALK5) in turn transduces a number of secondary signals, most notably the activation of Smad2/3. TβRI thus

* Correspondence: catherine.bauge@unicaen.fr; karim.boumediene@unicaen.fr
[1]Université Caen, IFR ICORE 146, Laboratory of Extracellular Matrix and Pathology, Esplanade de la Paix, 14032 Caen cedex, France
Full list of author information is available at the end of the article phosphorylates the receptor-regulated Smads (R-Smads) Smad2 and Smad3, which bind to Smad4, translocate into the nucleus and regulate gene expression in concert with other transcriptional factors, such as specific protein 1 (Sp1) [2,3]. Like R-Smads, the inhibitory Smad7 interacts with the activated type I TGFβ receptor. In contrast to Smad2/3, however, Smad7 forms a stable association with the receptor complex and prevents receptor-mediated phosphorylation of pathway-restricted Smads, resulting in disruption of TGFβ signalling [4].

In the cartilage context, it is thought that TGFβ signalling pathway plays a critical role for maintenance of tissue homeostasis, and modification of TGFβ signalling gene expression may be a cause for articular diseases such as osteoarthritis (OA) [5]. TβRII and Smad3, at least, are mediators of OA, as established using *in vitro* and *in vivo* models. Indeed, Smad3 gene mutations in humans or targeted disruption in mice are associated with the pathogenesis of OA [6,7]. Similarly, mice that express a cytoplasmically truncated type II receptor, which acts as a dominant-negative mutant, develop a degenerative joint disease resembling human OA [8]. In addition, *in vivo* OA is associated with modifications of TβRII and Smad7 expression [9,10].

Several studies reported that TGFβ levels are increased, at least in the first stage of the disease [1,9]. We therefore wondered whether the modifications of expression of TGFβ signalling mediators observed during OA may be due, in part, to a feedback loop of TGFβ.

Among numerous factors involved in the OA process and known to have the ability to regulate expression of TGFβ signalling genes, Sp1 seems to be particularly interesting. This protein is a trans-activator of cartilage-specific genes. The Sp1 knockdown is thus associated with reduction of collagen expression [11]. Sp1 is also involved in the regulation of Sox9 [12]. This transcriptional factor also cooperates with Smads to regulate expression of multiple TGFβ target genes [2,3,13].

In the present report, we have investigated the effect of TGFβ1 treatment on expression of TGFβ signalling genes (receptors and Smads) and downstream genes (Sox9, COL2A1, aggrecan, COL10A1, COL1A1) in human articular chondrocytes. We demonstrate that whereas TGFβ treatment upregulates its receptors and Smad3 after short exposition time of TGFβ1 (< 1 hour), it causes a dramatic decrease of both TGFβ receptors, and of Smad3 expression after longer incubation. In marked contrast, the levels of antagonistic Smad7 were increased in TGFβ-stimulated cells in all our experimental conditions. In addition, we showed that TGFβ1 induces a differential response according to the duration of treatment, with more beneficial effect for cartilage under short TGFβ exposition. We also established a role of Sp1 transcription factor in the downregulation of TGFβ receptors, and chondrocyte response to TGFβ. Taken together, these results provide novel insights for the auto-modulation of TGFβ signalling in chondrocytes.

## Materials and methods
### Reagents
Reagents were provided by Invitrogen (Bioblock Scientific, Illkirch, France) unless otherwise noted. TGFβ1 (R&D Systems, Lille, France) was resuspended in PBS-HCl. Mithramycin and actinomycin D were obtained from Sigma-Aldrich Co. (St Quentin Fallavier, France). Oligonucleotides were supplied by Eurogentec (Angers, France).

### Cell culture
OA human articular chondrocytes were prepared from femoral heads of patients who underwent hip replacement (ages between 63 and 81 years, median 77 years) as previously described [14]. All donors signed the agreement for this study according to the local ethical committee (Comité de protection des personnes). Cells were seeded at $4 \times 10^4$ cells/cm$^2$ and cultured in DMEM supplemented with 10% heat-inactivated FCS, 100 IU/ml penicillin, 100 μg/ml streptomycin and 0.25 μg/ml fungizone, in a 5% $CO_2$ atmosphere. Cells were cultured for 5 to 6 days in 10% FCS-containing DMEM. Then, at confluence, the cells were incubated in DMEM + 2% FCS for 24 hours before adding TGFβ1 (1 to 10 ng/ml) in the same medium.

### RNA extraction and real-time RT-PCR
Total RNA from primary human articular chondrocyte cultures was extracted using Trizol. Following extraction, 1 μg DNase-I treated RNA was reverse transcribed into cDNA as previously described [14]. Amplification of the generated cDNA was performed by real-time PCR in Applied Biosystems SDS7000 apparatus (Applied Biosystems Inc., Courtaboeuf, France). The relative mRNA level was calculated with the $2^{-\Delta\Delta Ct}$ method. Primer sequences are presented in Table 1.

### Protein extraction and western blot analysis
Cells were rinsed, and scrapped in RIPA lysis buffer supplemented with phosphatase and protease inhibitors. The extracts (50 μg protein) were subjected to fractionation in 10% SDS-PAGE, transferred to polyvinylidene fluoride membranes (Amersham Biosciences, Orsay, France), and reacted with TβRI, TβRII, Smad2/3 or phospho-Smad2/3 polyclonal antibodies (Tebu-bio, Le Perray en Yvelines, France). Subsequently, membranes were incubated with appropriate secondary peroxidase-conjugated antibody. The signals were revealed with SuperSignal West Pico Chemiluminescent Substrate (Pierce Perbio Science, Brébières, France) and exposed to X-ray film. The membranes were also reacted with anti β-actin to verify equal loading.

Table 1 Primer sequences for the present study

| Primer | Sequence (5' to 3') | |
|---|---|---|
| | Sense | Antisense |
| TβRI | TTAAAAGGCGCAACCAAGAAC | GTGGTGATGAGCCCTTCGAT |
| TβRII | GACATCAATCTGAAGCATGAGAACA | GGCGGTGATCAGCCAGTATT |
| Smad2 | GCTGTTTTCCTAGCGTGGCTT | TCCAGACCCACCAGCTGACT |
| Smad3 | GCATCAGCCGCTTCTCAAGT | ATCTCCCCACCATCACCTCC |
| Smad4 | CCTTCTGGAGGAGATCGCT | TCAATGGCTTCTGTCCTGTGG |
| Smad7 | AATGTGTTTTCTAGATTCCCAACTTCTT | CACTCTCGTCTTCTCCTCCCAGTA |
| Sp1 | AGAATTGAGTCACCCAATGAGAACA | GTTGTGTGGCTGTGAGGTCAAG |
| COL2A1 | GGCAATAGCAGGTTCACGTACA | CGATAACAGTCTTGCCCCACTT |
| COL1A1 | CACCAATCACCTGCGGTACAGAA | CAGATCACGTCATCGCACAAC |
| COL10A1 | CCTGGTATGAATGGACAGAAAGG | CCCTGAGGGCCTGGAAGA |
| Aggrecan | TCGAGGACAGCGAGGCC | TCGAGGGTGTAGCGTGTAGAGA |
| Sox9 | CCC ATG TGG AAG GCA GAT G | TTC TGA GAG GCA CAG GTG ACA |

### Transfection experiments
Sp1 expression vector (pEVR2-Sp1) was obtained from Dr Suske (Institut fur Molekularbiologie and tumorforschung, Marburg, Germany). Chondrocytes were transiently transfected by the nucleofection method as previously described [14]. After overnight transfection, cells were treated with TGFβ1 (5 ng/ml) in DMEM containing 2% FCS. The silencing of Sp1 was performed using a siRNA targeting Sp1 (Tebu-Bio; Sp1 siRNA (h), sc-29487: AAUGAGAACAGCAACAACUCC) or a control sequence (UUGUCCGAACGUGUCACGUdtdt), as previously described [13].

### Statistical analysis
All experiments were repeated with different donors at least three times with similar results, and representative experiments are shown in the figures. Data are presented as the mean ± standard deviation. Statistical significance was determined by Student's $t$ test. Differences were considered statistically significant at $P < 0.05$.

## Results
### TGFβ1 downregulates TGFβ receptors and Smad3, and upregulates Smad7
We investigated the effect of TGFβ1 on mRNA expression of TGFβ signalling genes in a dose-dependent manner, using real-time RT-PCR (Figure 1). A 48-hour incubation with TGFβ1 significantly reduced the expression of both TGFβ receptors and Smad3, whereas the Smad7 mRNA level was increased. These effects were maximal at 1 ng/ml, except for TβRII for which the maximal effect was observed only at doses above 5 ng/ml. No significant effect was observed on Smad2 and Smad4.

### TGFβ1 differentially regulates expression of its receptors and Smad3 according to duration of incubation
A time-course study (Figure 2a) revealed that, at mRNA levels, TGFβ1 quickly upregulates its own receptors and Smad3, since it increases their expression as soon as 30 minutes of treatment. For longer treatments, TGFβ1 exerted the opposite effect and downregulated TGFβ receptors (after 24 hours of incubation) as well as Smad3 (after 3 hours of incubation). On the contrary, TGFβ1 upregulated Smad7 expression whatever the time of incubation.

Furthermore, western blot analysis (Figure 2b) showed that TβRII is downregulated after 24 hours whereas TβRI protein expression is decreased as soon as 1 hour after TGFβ1 treatment. In addition, as expected, TGFβ1 induced Smad2/3 phosphorylation - but this effect is transient since we were no longer able to detect phosphorylated Smad2/3 after 3 hours or 24 hours of treatment with TGFβ1.

### TGFβ exerts differential effects on matrix genes and Sox9 according to duration of treatment
To evaluate the importance of the regulation of TGFβ pathways in cartilage homeostasis, we analysed mRNA expression of matrix genes (collagens type II, type I, and type X, and aggrecan) after increased duration of treatment (from 1 to 48 hours) (Figure 3). TGFβ1 acted with various kinetics according to the considered genes. It induced COL2A1 expression in a biphasic manner (at 3 hours and after 24 hours of treatment, with no stimulation for 6 hours of incubation). TGFβ1 repressed aggrecan expression after 6 hours of treatment, and upregulated COL1A1 as soon as 1 hour of incubation.

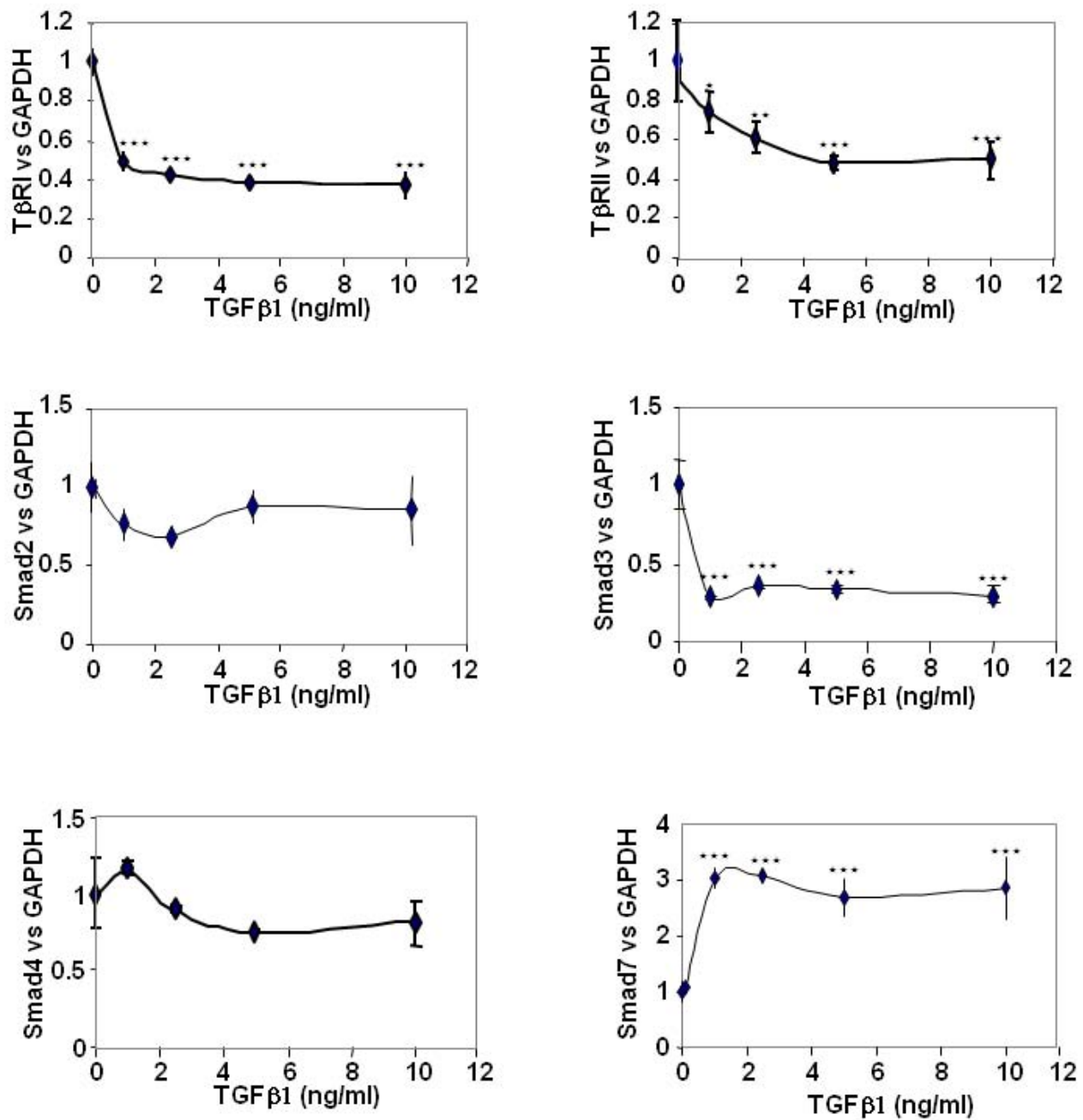

Figure 1 Transforming growth factor beta 1 (TGFβ1) downregulates TGFβ receptors and Smad3, and upregulates Smad7. Human articular chondrocytes (HAC) were cultured for 5 to 6 days in 10% FCS-containing DMEM. They were then incubated in DMEM + 2% FCS with increasing doses of transforming growth factor beta 1 (TGFβ1) for 48 hours. TGFβ receptor type I (TβRI), TGFβ receptor type II (TβRII), Smad2, Smad3, Smad4 and Smad7 mRNA were analysed by real-time RT-PCR. The modulation of mRNA expression was expressed relative to the controls (not treated), after normalisation to the GAPDH signal. *, $P < 0.05$, **, $P < 0.01$, ***, $P < 0.001$.

Concerning hypertrophic markers of cartilage, TGFβ1 induced collagen type X expression after 24 hours of incubation. We also focused our attention on Sox9, a major transcription factor for the chondrocyte phenotype, and found that TGFβ1 induced its expression only for 1 hour of incubation.

### TGFβ1 enhances TGFβ receptor mRNA turnover, but does not modify that of Smads

Modifications of gene expression under TGFβ treatment could be due to an increased degradation rate and/or a reduced transcription. We therefore asked whether TGFβ1 affects mRNA decay of TβRI, TβRII, Smad3 and Smad7. Human articular chondrocytes were incubated with actinomycin D, a transcription inhibitor, in addition to TGFβ (Figure 4). The half-lives of Smad3 and Smad7 mRNA, which were approximately 3.5 hours and 45 minutes, respectively, were not significantly modified by TGFβ. On the contrary, inhibition of *de novo* transcription clearly showed that TGFβ reduced the mRNA half-life of both TGFβ receptors. Indeed, the TβRI half-life is about 20 minutes but was reduced to 10 minutes when chondrocytes were incubated with TGFβ, and the TβRII mRNA half-life is 45 minutes for control cells and was reduced by almost 80% after TGFβ treatment.

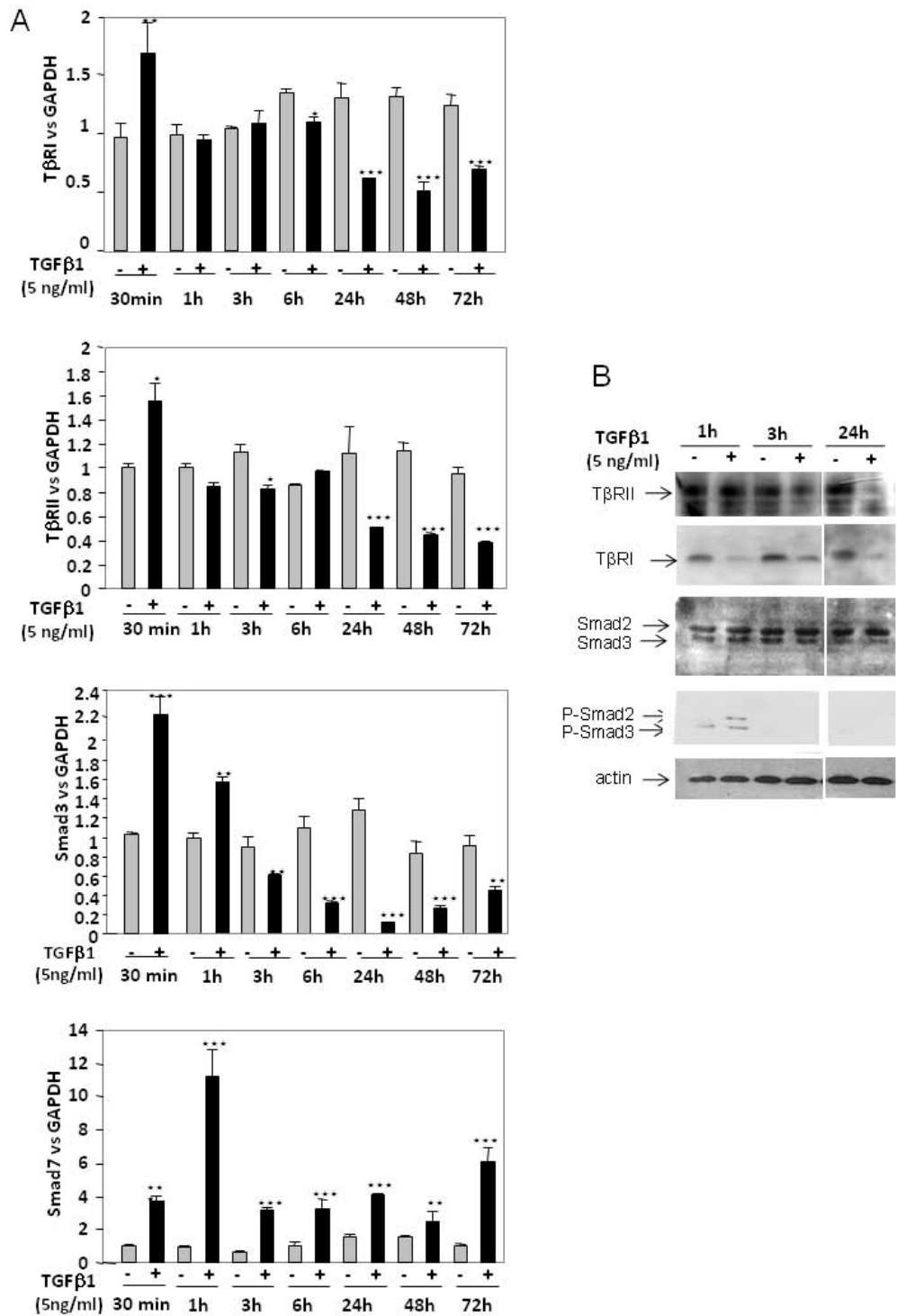

**Figure 2 Transforming growth factor beta 1 regulation of receptors and Smad3 expression according to incubation duration.**
**(a)** Human articular chondrocytes (HAC) were cultured as in Figure 1 and incubated with 5 ng/ml transforming growth factor beta 1 (TGFβ1) for different times. At the end of incubations, TGFβ receptor type I (TβRI), TGFβ receptor type II (TβRII), Smad3 and Smad7 mRNA levels were assayed by real-time RT-PCR. **(b)** In addition, TβRI, TβRII, Smad2/3 and phosphorylated Smad2/3 protein expression were analysed by western blot analysis. *, $P < 0.05$, **, $P < 0.01$, ***, $P < 0.001$.

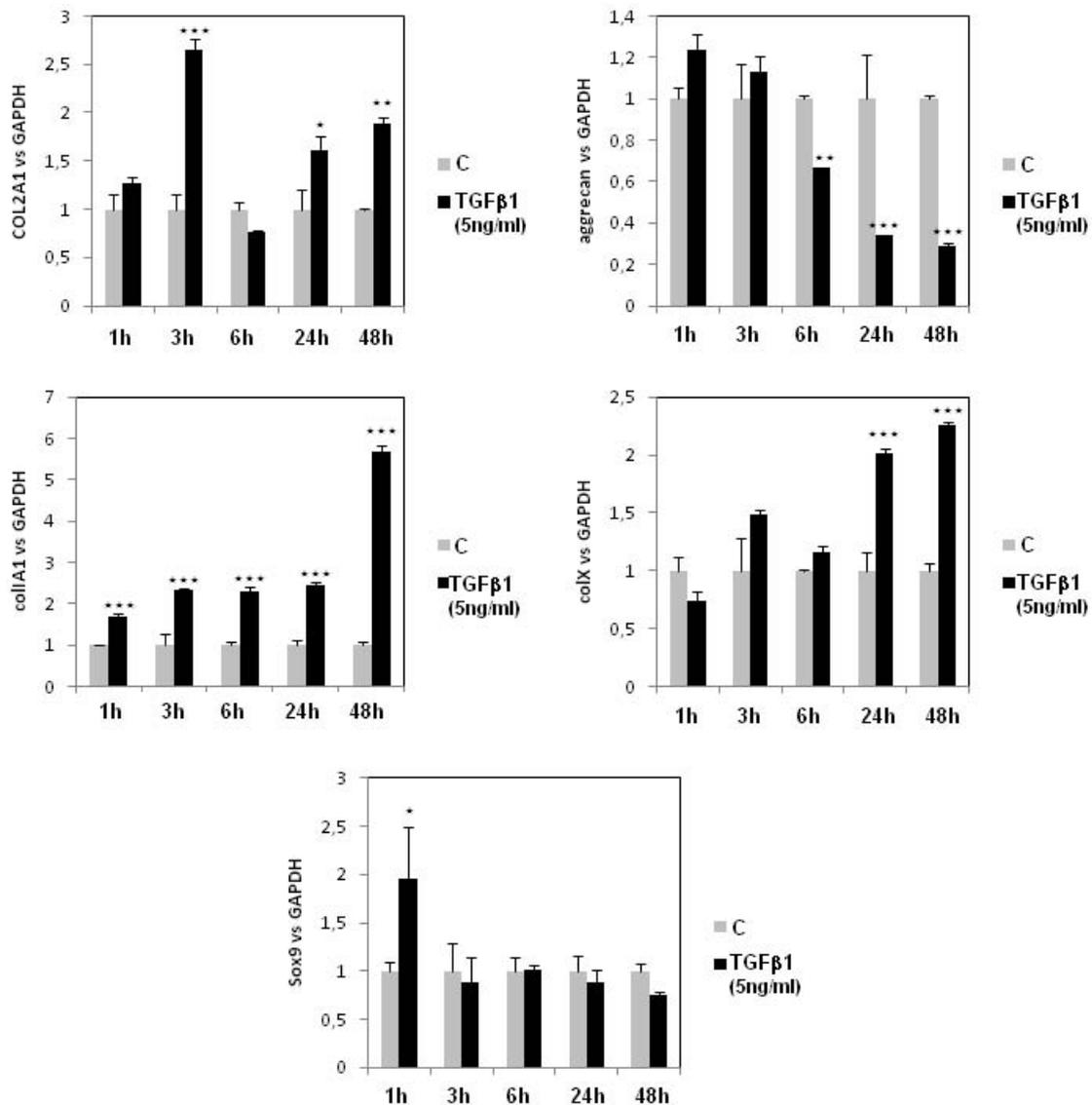

Figure 3 Transforming growth factor beta differential effects on matrix genes and Sox9 according to treatment duration. Human articular chondrocytes (HAC) were cultured and incubated as Figure 2. COL1A1, COL2A1, COL10A1, aggrecan and Sox9 mRNA levels were then determined by RT-PCR. C, control. *, $P < 0.05$, **, $P < 0.01$, ***, $P < 0.001$.

### Sp1 mediates TGFβ-induced modulation of TGFβ receptors

As mentioned above, Sp1 is important for cartilage metabolism. We therefore analysed the effect of TGFβ1 on Sp1 expression. We showed that TGFβ strongly reduces Sp1 mRNA levels in a dose-dependent and time-dependent manner (Figure 5).

To further investigate the putative role of Sp1, TGFβ signalling gene expression was analysed in the presence of mithramycin, an inhibitor of DNA binding of Sp1 family members. Inhibition of Sp1 binding for 24 hours mimics TGFβ-induced repression of receptor expression, whereas it does not affect Smad expression (Figure 6a).

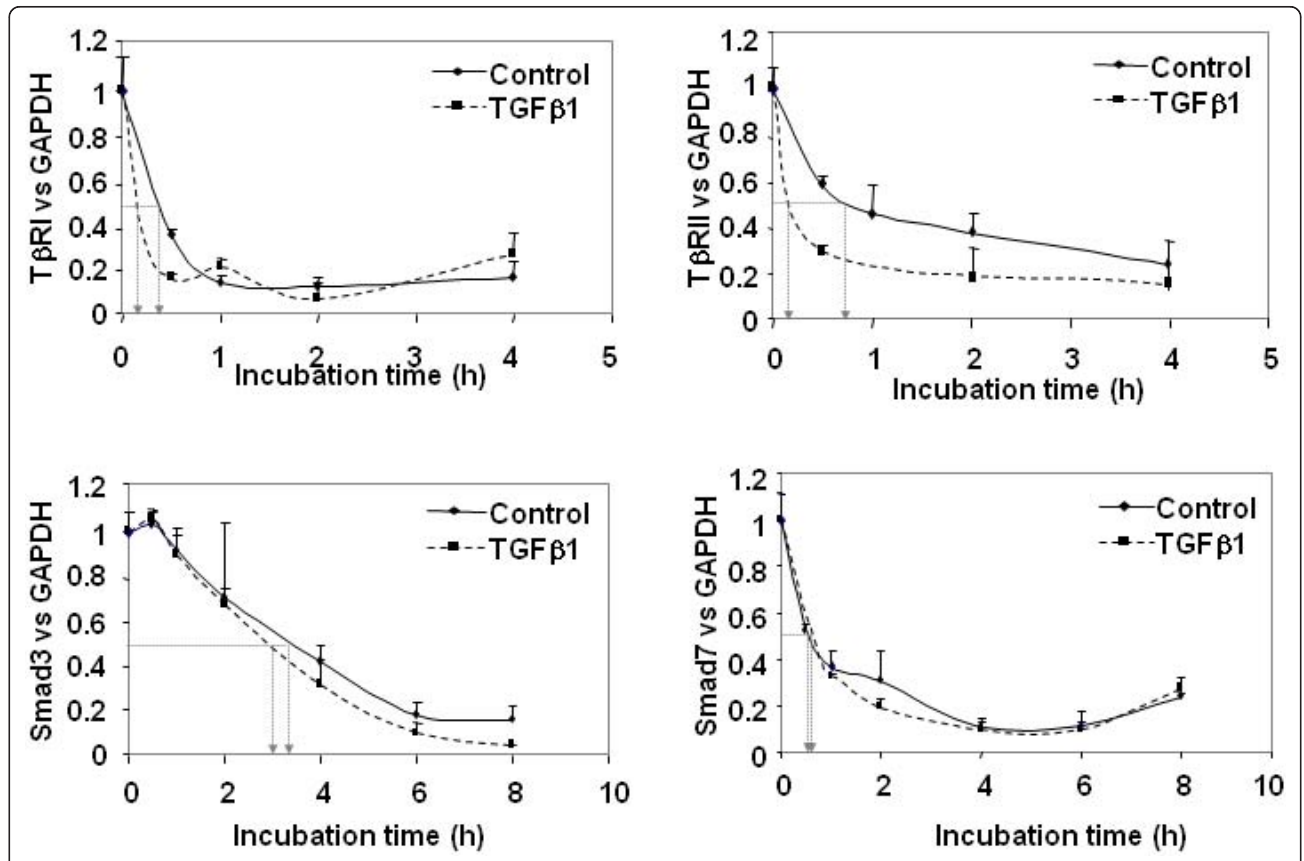

Figure 4 Transforming growth factor beta 1 (TGFβ1) enhances TGFβ receptors mRNA turnover. Subconfluent human articular chondrocytes (HAC) were incubated with DMEM + 2% FCS for 24 hours. Thereafter, transforming growth factor beta 1 (TGFβ1) or vehicle were added in the presence of actinomycin D (10 μg/ml). Cells were then harvested at the indicated times for RT-PCR.

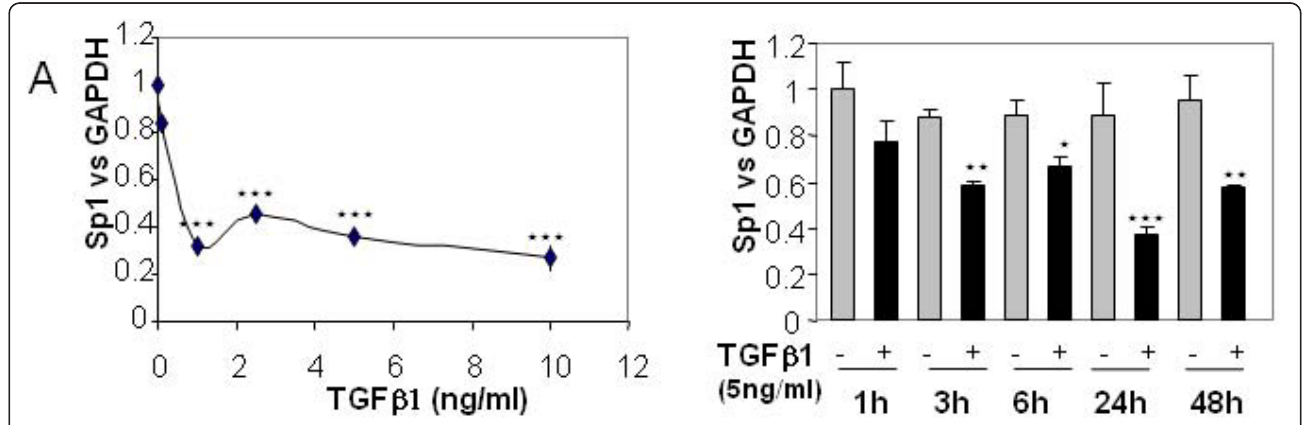

Figure 5 Transforming growth factor beta reduces specific protein 1 mRNA levels dose and time dependently. Human articular chondrocytes (HAC) were cultured for 5 to 6 days in 10% FCS-containing DMEM. They were then incubated in DMEM + 2% FCS for 24 hours, before addition of increased concentrations of transforming growth factor beta 1 (TGFβ1) or vehicle. mRNA levels of specific protein 1 (Sp1) were analysed by real time RT-PCR. HAC were also treated with 5 ng/ml TGFβ and the Sp1 mRNA level was determined. *, $P < 0.05$, **, $P < 0.01$, ***, $P < 0.001$.

To confirm the specific role of Sp1 in these regulations, gain and loss of function experiments were performed. First, silencing of Sp1 by siRNA for 24 hours led to inhibition of both TGFβ receptor expression but did not modify Smad3 and Smad7 expression (Figure 6b). In contrast, forced expression of Sp1 for 24 hours did not change TβRI and TβRII expression but counteracted TGFβ-induced repression on these genes, whereas it did not affect Smad expression either in the presence or in the absence of TGFβ (Figure 6c). The depletion of Sp1 by siRNA and the overexpression of Sp1 in pEVR2-Sp1 transfected cells were checked by western blot analysis (Figure 7) [13].

### Sp1 ectopic expression permits maintaining a similar expression pattern as early response to TGFβ even after 24 hours of treatment

Since ectopic expression of Sp1 permits one to counteract the inhibition of TβRI and TβRII expression induced by long treatment with TGFβ, we hypothesised that it may also affect the expression of downstream genes. We therefore investigated the expression of matrix genes after 24 hours of incubation with TGFβ1 in cells that had been transfected with Sp1 expression vector or control vector. Ectopic expression of Sp1 modified cell responses to TGFβ. In Sp1 transfected chondrocytes, 24-hour treatment with TGFβ induced COL2A1 and Sox9 upregulation but was not able to downregulate aggrecan. Additionally, Sp1 ectopic expression blocked the upregulation of COL10A1 and COL1A1. Interestingly, the gene expression pattern induced by TGFβ1 at 24 hours under Sp1 ectopic expression (Figure 8) is similar to the early effect of TGFβ1 at 1 hour in untransfected cells (Figure 3).

### Discussion

To our knowledge, the present study is the first systematic analysis of regulation by TGFβ on gene expression of its own receptors and Smads, in human articular chondrocytes. Our study shows that TGFβ exerts a differential effect on the transcription of genes implicated in the canonical Smads pathway. While TGFβ upregulates its receptors and Smad3 for short incubation (at least at mRNA level), it downregulates them in the long term. In addition, it upregulates Smad7 and does not significantly alter Smad2 and Smad4 expression. This positive and negative feedback loop of the TGFβ pathway induces differential response of chondrocytes to TGFβ. The mechanisms responsible for modulation of Smads and for TGFβ receptor expression seem to be different. Indeed, TGFβ downregulates both receptors, at least by modifying the mRNA stability. This process appears slowly (after 24 hours of treatment). On the contrary, TGFβ1 quickly regulates Smad3 and Smad7 mRNA levels by a mechanism independent of mRNA stability.

Our results suggest that following TGFβ1 administration a rapid activation of TGFβ signalling occurs, characterised by phosphorylation of Smad2/3 and upregulation of TβRI, TβRII and Smad3 (at least at mRNA level). Thereafter, a negative feedback loop of the TGFβ1 signalling pathway occurs with a decline of these receptors and R-Smad expression and a simultaneous rise in the inhibitory Smad7 level. The activation of P-Smad2/3 and upregulation of Smad7 after 30 minutes of TGFβ treatment are consistent with observations from Jimenez's group obtained with human and bovine chondrocytes [15].

The downregulation of TGFβ receptors by its own ligand is controversial, and is dependent on cell type as well as on duration of TGFβ1 incubation. In lung fibroblasts, TGFβ1 induced an increased type I receptor expression by enhancing the transcription of this gene [16], whereas its expression is not modulated or downregulated in osteoblasts [17,18]. Similarly, TβRII can be downregulated or upregulated by its own ligand [18-20]. In addition, in osteoblasts TGFβ1 reduces the amount of specific TβRII at the cell surface but does not affect the mRNA steady-state level [21].

We have established that, in human OA chondrocytes, TGFβ acts, at least in part, by strongly decreasing the mRNA stability of its receptors. This rapid turnover potentially allows the receptor rate to change rapidly in response to its own ligand. We cannot, however, exclude the possibility that TGFβ downregulates its receptors also at the transcriptional and translational levels.

Concerning Smad effectors, our results are consistent with data obtained in normal skin fibroblasts [22] - which demonstrated that TGFβ treatment causes an upregulation of antagonistic Smad7, and a dramatic decrease in Smad3 mRNA expression. Interestingly, the mRNA level of the closely related Smad2 was not affected by 48 hours of treatment with TGFβ1. A differential regulation between R-Smads has already been described in lung epithelial and mesangial cells [23,24] and may lead to a variation in the cell response according to the level of TGFβ. Similar to findings obtained in fibroblasts [22] or in mesangial cells [24], we established that the downregulation of Smad3 mRNA expression in TGFβ-treated chondrocytes was not due to decreased transcript stability, suggesting a transcriptional effect of TGFβ. Further experiments, such as nuclear run-on or gene reporter assays, would be required to definitively state this hypothesis.

In contrast to Smad3, Smad7 mRNA expression was rapidly and markedly induced by TGFβ. These findings are agreement with reports describing Smad7 as an immediate-early gene target of TGFβ in MV1Lu cells, HaCaT cells [4] and skin fibroblasts [22]. Increased expression of the inhibitor Smad7 has been associated with inhibition of TGFβ signalling. Smad7 could negatively regulate TGFβ signalling; on one hand by inhibiting R-Smad activation by TβRI or by enhancing TβRI degradation in the cytoplasm, and on the other hand by disrupting the formation of the TGFβ-induced functional Smad-DNA complex in the nucleus [25].

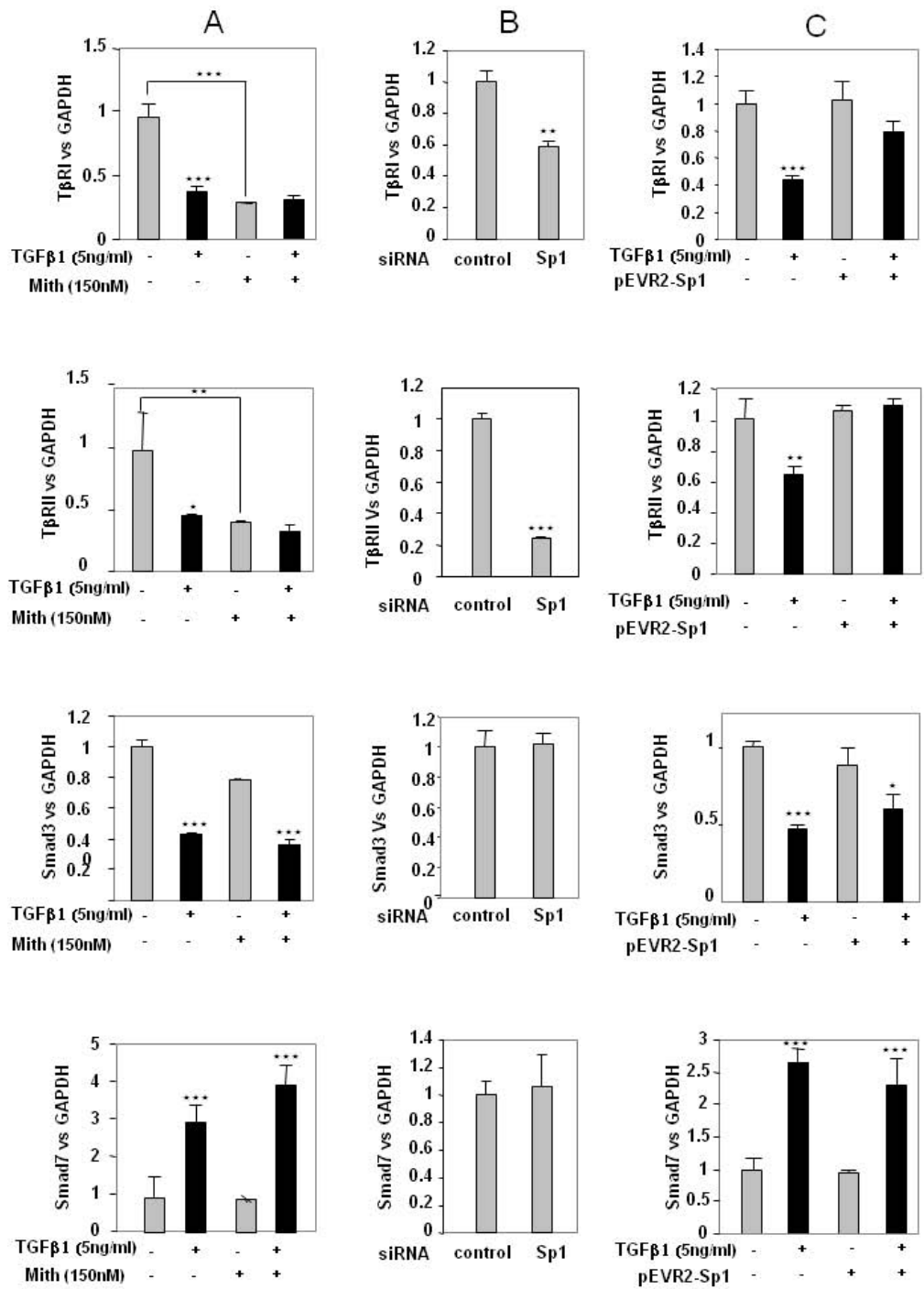

Figure 6 Specific protein 1 mediates transforming growth factor beta (TGFβ)-induced modulation of TGFβ receptors. (a) Subconfluent cultures of chondrocytes were treated for 24 hours in the presence or absence of mithramycin (150 nM). TGFβ receptor type I (TβRI), TGFβ receptor type II (TβRII), Smad3 and Smad7 expression was analysed at the mRNA level by real-time RT-PCR. (b) Human articular chondrocytes (HAC) were also nucleofected with specific protein 1 (Sp1) siRNA oligonucleotides or control sequence. Thereafter, the medium was replaced with DMEM + 10% FCS for 24 hours. Total RNA was then extracted and real-time RT-PCR analysis was performed. Histograms represent the relative TβRI, TβRII, Smad3 or Smad7 mRNA levels versus GAPDH. (c) HAC were transfected overnight with pEVR2-Sp1 (or with insertless plasmid as controls). Thereafter, media were replaced with DMEM + 2% FCS for 24 hours in the absence or the presence of transforming growth factor beta 1 (TGFβ1) (5 ng/ml). Therefore, TβRI, TβRII, Smad3 or Smad7 mRNA levels were analysed and expressed as relative expression versus GAPDH. *, $P < 0.05$, **, $P < 0.01$, ***, $P < 0.001$.

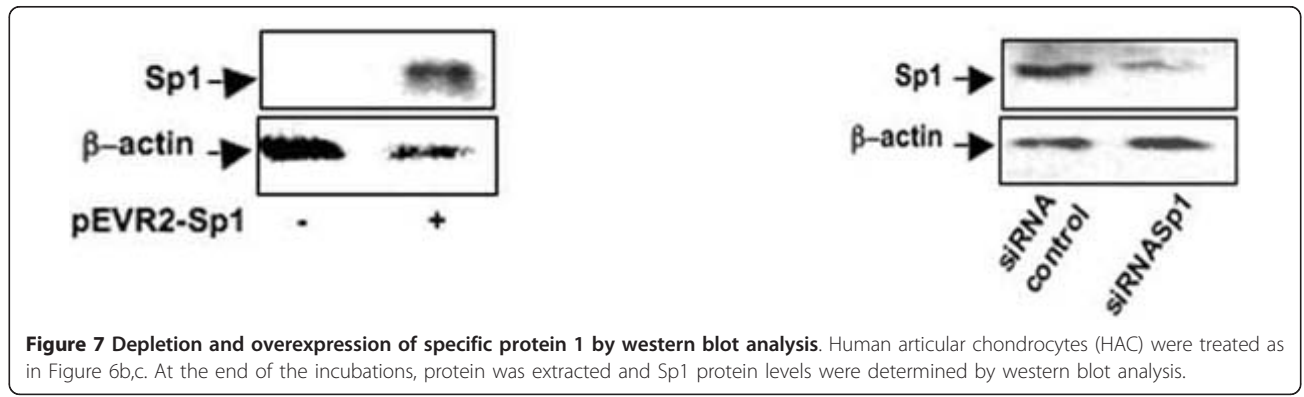

**Figure 7 Depletion and overexpression of specific protein 1 by western blot analysis**. Human articular chondrocytes (HAC) were treated as in Figure 6b,c. At the end of the incubations, protein was extracted and Sp1 protein levels were determined by western blot analysis.

These TGFβ-induced modifications on expression of TGF receptors and Smads may participate in the chondrocyte-phenotype changes observed in OA, a pathology associated, at least in the first stage, with an increase in the TGFβ level [9]. Modifications of Smad3 expression are associated with OA [6,7], and its expression stimulates type II collagen synthesis caused by TGFβ1 [26]. Moreover, activation of Smad pathways by transfection with a dominant-negative Smad7 retroviral vector or constitutively active TβRII abolished retinoic acid-induced inhibition of chondrogenesis, suggesting that TGFβ receptor/Smad signalling is essential for this process [27]. Furthermore, ectopic expression of TβRII restores TGFβ sensitivity and increases aggrecan and col2 expression, in IL1-treated or passaged chondrocytes, respectively ([14] and unpublished personal data).

Our experiments indicate that TGFβ1 exerts a differential effect on profiling of gene expression in chondrocytes according to the duration of treatment. A short TGFβ1 administration (1 hour) induces Sox9 expression, followed, after 3 hours, by induction of collagen type II expression. This effect was transient, but a second peak of collagen II expression appears after 24 hours of incubation of TGFβ1. These data suggest that at least two different mechanisms are responsible for cell response to TGFβ. A short TGFβ administration may activate the Smad2/3 pathway (upregulation of TβRI, TβRII and Smad3, and phosphorylation of Smad2/3), leading to an increase of Sox9, which, in turn, may induce collagen type II expression. Thereafter, a negative feedback loop occurs, characterised by a reduction of TβRI, TβRII and Smad3 expression and simultaneous induction of the inhibitory Smad7. This feedback leads to blockage of Smad2/3-mediated TGFβ signalling and reduction of Sox9, and furthermore to reduced collagen type II expression.

On the contrary, longer incubation leads an additional response to TGFβ but with a different pattern of matrix gene expression. This late response is associated with increased atypical collagen expression (COL1A1 and COL10A1) and reduction of aggrecan expression. These data suggest that a noncanonical pathway could be involved in this late response to TGFβ. Several pathways may be implied. In particular, the reduction of TβRI expression may change the ratio between TβRI and ALK1, another type I TGFβ receptor recently identified in chondrocytes, favouring TGFβ signalling via the Smad1/5/8 route and, subsequently, chondrocyte terminal differentiation [28,29].

Finally, in the present report we show that Sp1 is involved in the regulation of TGFβ receptors and cell response to TGFβ. TGFβ acts controversially on Sp1 expression. Previous data obtained in rabbit chondrocytes showed that TGFβ decreases Sp1 expression and binding activity [30], whereas recent studies indicate that TGFβ induces Sp1 in skin fibroblasts [31]. Our data show that Sp1 is downregulated in human chondrocytes, suggesting that this negative effect does not depend on the species but is cell-type specific.

The mechanism by which TGFβ regulates Sp1 expression is still unclear. In particular, the role of Smads in the regulation of Sp1 promoter activity is not known. Analysis of the Sp1 promoter (region -2,000 to +1) with Patch_Search [32], however, shows numerous putative binding sites for Smad3 and Smad4 in the 1,000 base pair upstream transcription initiation site of the Sp1 gene. An extensive study will be required to determine whether Smads directly or indirectly regulate Sp1 expression. Besides, a recent study shows that Smads bind in association with Sp1 to the CC(GG)-rich TGFβ1 responsive element of the human α1 type I collagen promoter that lacks the classical Smad recognition element, thus enhancing the binding of Sp1 and in this manner activating the collagen promoter [33]. Numerous studies indicate also that Sp1 cooperate with Smads to regulate the expression of TGFβ target genes [3,31,34,35].

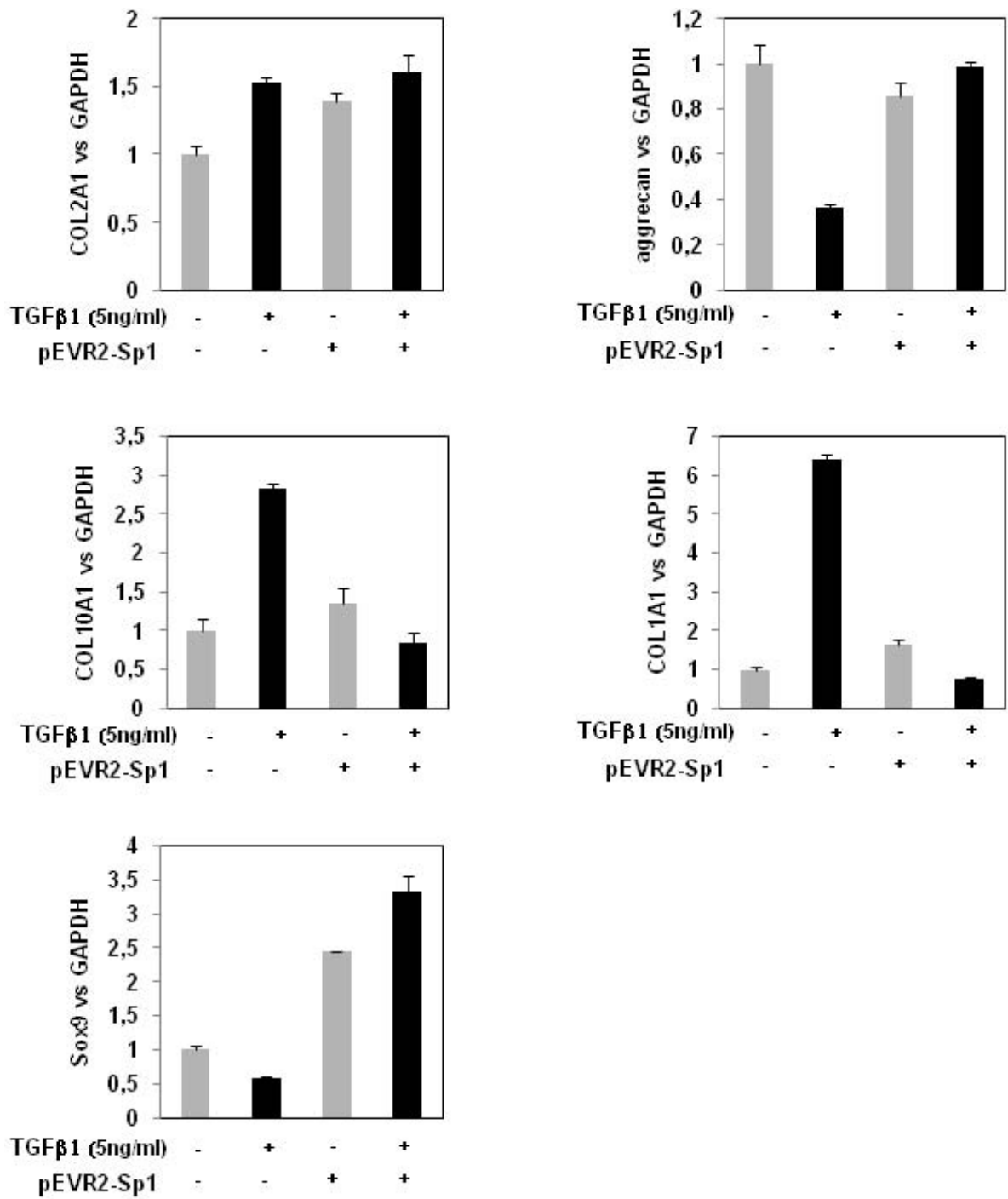

**Figure 8 Gene expression pattern induced by transforming growth factor beta 1 under Sp1 ectopic expression**. Human articular chondrocytes (HAC) were nucleofected with pEVR2 or pEVR2-Sp1. Thereafter, the media were replaced with DMEM + 2% FCS for 24 hours before adding transforming growth factor beta 1 (TGFβ1) (5 ng/ml) for an additional 24 hours of incubation. Total RNA was then extracted and real-time RT-PCR analysis was performed. Histograms represent the relative COL1A1, COL2A1, COL10A1, aggrecan or Sox9 mRNA levels versus GAPDH.

Importantly, restoration by Sp1 of TGFβ receptor expression after inhibition by TGFβ1 strongly suggests that inhibition of Sp1 by TGFβ is a potential cause of TGFβ-mediated suppression. These results were in agreement with previous reports that demonstrate Sp1 is a transactivator of both TGFβ receptors [36,37]. Moreover, a key role of Sp1 in the Smad7 induction by TGFβ was recently established in pancreatic cancer cells [3]. In our study, however, Sp1 does not regulate Smad7 expression, suggesting that the regulatory mechanism of Smad7 is cell specific.

Interestingly, Sp1 ectopic expression permits one to maintain, even after 24 hours of treatment, the early cell response to TGFβ (induction of Sox9, COL2A1) and to counteract the late response (upregulation of COL1A1, COL10A1, repression of aggrecan). These data suggest that targeting Sp1 expression in association to TGFβ treatment might be an innovative strategy to maintain or induce the chondrocyte phenotype.

## Conclusions

The present study enlightens a mechanism of feedback loop controlling TGFβ responses in human OA chondrocytes. Contrary to previous studies, which examined one particular gene, we investigated the TGFβ-induced expression of both TGFβ receptors and Smads, and the molecular mechanism involved. We show that brief administration of TGFβ induces its signalling with upregulation of TGFβ receptors and Smad3, which is associated with Sox9 and COL2A1 induction. On the contrary, a long incubation with TGFβ downregulates its own receptors by decreasing the mRNA stability, reduces the Smad3 expression and upregulates the inhibitor Smad7. In addition, long treatments do not induce Sox9 expression but upregulate atypical cartilage matrix genes such as COL1A1 and COL10A1. We also provide information about the mechanism involved in this regulation. We showed the implication of the transcriptional factor Sp1 in the repression of both TGFβ receptors but not in the modulation of Smad3 and Smad7. In addition, we demonstrated the involvement of Sp1 in both early and late response of these cells to TGFβ. Sp1 ectopic expression permitted one to maintain the early response of OA chondrocytes to TGFβ at 24 hours of treatment. Together, these data provide an overall view of the feedback loop of the TGFβ signal in human articular chondrocytes, and highlight an interesting role of Sp1 in regulating the TGFβ response.

### Abbreviations
DMEM: Dulbecco's modified Eagle's medium; FCS: foetal calf serum; OA: osteoarthritis; PBS: phosphate-buffered saline; PCR: polymerase chain reaction; R-Smads: receptor-regulated Smads; RT: reverse transcriptase; siRNA: small interfering RNA; Sp1: specific protein 1; TβRI: TGFβ receptor type I; TβRII: TGFβ receptor type II; TGFβ: transforming growth factor beta.

### Acknowledgements
The authors thank Dr Suske (Institut fur Molekularbiologie and tumorforschung, Marburg, Germany) for providing pEVR2-Sp1. OC is a recipient of a fellowship from the Conseil Régional de Basse-Normandie.

### Author details
[1]Université Caen, IFR ICORE 146, Laboratory of Extracellular Matrix and Pathology, Esplanade de la Paix, 14032 Caen cedex, France. [2]Department of Orthopaedic Surgery, Saint-Martin Private Clinic, Rue Roquemonts, 14000 Caen, France.

### Authors' contributions
CB conceived and carried out experiments, analysed data and wrote the paper. OC and SL participated in data collection and analysis. PG participated in data interpretation. KB conceived experiments, carried out experiments and analysed data. All authors were involved in writing the paper and had final approval of the submitted and published versions

### Competing interests
The authors declare that they have no competing interests.